\documentclass[%
 reprint,
superscriptaddress,
 amsmath,amssymb,
 aps,
]{revtex4-2}

\usepackage{graphicx}
\usepackage{dcolumn}
\usepackage{bm}
\usepackage{color}

\newcommand{\addihpc}{Institute of High Performance Computing, A$^\star$STAR (Agency for Science, Technology and Research), Singapore}

\newcommand{\addeinhoven}{Department of Applied Physics and Science Education, Eindhoven University of Technology, The Netherlands}

\newcommand{\addsutd}{Science, Mathematics and Technology Cluster, Singapore University
of Technology and Design, Singapore}

\newcommand{\addanu}{Nonlinear Physics Center, Research School of Physics, Australian National University, Australia}

\begin{document}

\preprint{APS/123-QED}

\title{Unconventional localization of light with Mie-tronics}

\author{Thanh Xuan Hoang}
\email{hoangxuan11@gmail.com}
\affiliation{\addihpc}%

\author{Daniel Leykam}%
\affiliation{\addsutd}%

\author{Ayan Nussupbekov}
\affiliation{\addihpc}%

\author{Jie Ji}
\affiliation{\addeinhoven}%

\author{Jaime G\'omez Rivas}
\affiliation{\addeinhoven}

\author{Yuri Kivshar}%
\affiliation{\addanu}%

\date{\today}

\begin{abstract}
Localization of light requires high-$Q$ cavities or spatial disorder, yet the wave nature of light may open novel opportunities. Here we suggest to employ {\it Mie-tronics} as a powerful approach to achieve the hybridization of different resonances for the enhanced confinement of light via interference effects. Contrary to a conventional approach, we employ the symmetry breaking in finite arrays of resonators to boost the $Q$ factors by in-plane multiple scattering. Being applied to photonic moir\'e structures, our approach yields a giant enhancement of the Purcell factor via twist-induced coupling between degenerate collective modes. Our findings reveal how finely tuned cooperative scattering can surpass conventional limits, advancing the control of wave localization in many subwavelength systems.
\end{abstract}

\maketitle

{\it Introduction.---}Wave propagation and localization have long fascinated scientists, shaping our understanding of natural phenomena and enabling technologies across classical optics, acoustics, and quantum systems. Even the simplest case—such as scattering of waves by point scatterers and spheres—have inspired theoretical advances from the pre-Maxwell era to the modern rise of Mie-tronics~\cite{Logan1965survey, Kivshar2022rise}. It is now widely recognized that wave localization arises from coherent multiple scattering and interference of several waves~\cite{John1991localization}.

Conventional mechanisms for localization of waves fall into two broad categories: cavity-based and disorder-induced localization. Cavity-based modes—including whispering-gallery, Fabry-P\'erot (FP), and bandgap modes—originate in diverse physical domains, but they have converged conceptually under the shared language of wave physics. Despite their differences, they all rely on reflection-based confinement: internal reflection along curved boundaries, mirror reflection between planar interfaces, or Bragg reflection in periodic lattices~\cite{Vahala2003optical}.

Disorder-induced (or Anderson) localization, by contrast, is rooted in multiple electron scattering~\cite{Anderson1958absence}, with its optical analogue proposed through a connection between atomic orbitals and Mie resonances~\cite{John1987strong,Lagendijk1996resonant}. Though several experiments have claimed to observe optical analogues of Anderson localization~\cite{Wiersma1997localization,Hsieh2015photon}, recent advances in computational electromagnetism and Mie-tronics suggest that the radiative and long-range interaction nature of light may fundamentally prohibit true Anderson localization~\cite{Yamilov2023anderson,Hoang2025collective}. In fact, spatial disorder often degrades rather than enhances optical confinement, challenging long-standing beliefs and motivating the development of new strategies for efficient localization of light. 

Grounded in modern nanophotonics and rigorous electromagnetic theory, Mie-tronics offers a scattering-based framework built on multipolar expansions. Beyond providing a first-principles interpretation of cavity resonances, it unifies diverse localization mechanisms through coherent multipolar interactions.

Here and in the accompanying paper~\cite{Hoang2025symmetry}, we propose an unconventional route to light localization by combining key features of FP, whispering-gallery, and band-edge resonances. Within the Mie-tronics framework, we show that symmetry breaking in finite arrays—traditionally thought to degrade the $Q$ factor—can instead enhance in-plane multiple scattering, thereby increasing both $Q$ and Purcell factors. We apply this approach to photonic moir\'e structures, demonstrating that symmetry breaking lifts degeneracy and enables strong coupling between collective modes, resulting in tight confinement and giant Purcell enhancement. These results establish a general strategy for achieving high-$Q$ localization through cooperative interference, clarify the origins of flat bands and magic angles in moir\'e systems~\cite{Tarnopolsky2019origin,Wang2022intrinsic,Yu2023origin}, and provide a coherent alternative to disorder-based localization.

\begin{figure*}
\includegraphics[width = 1.8\columnwidth]{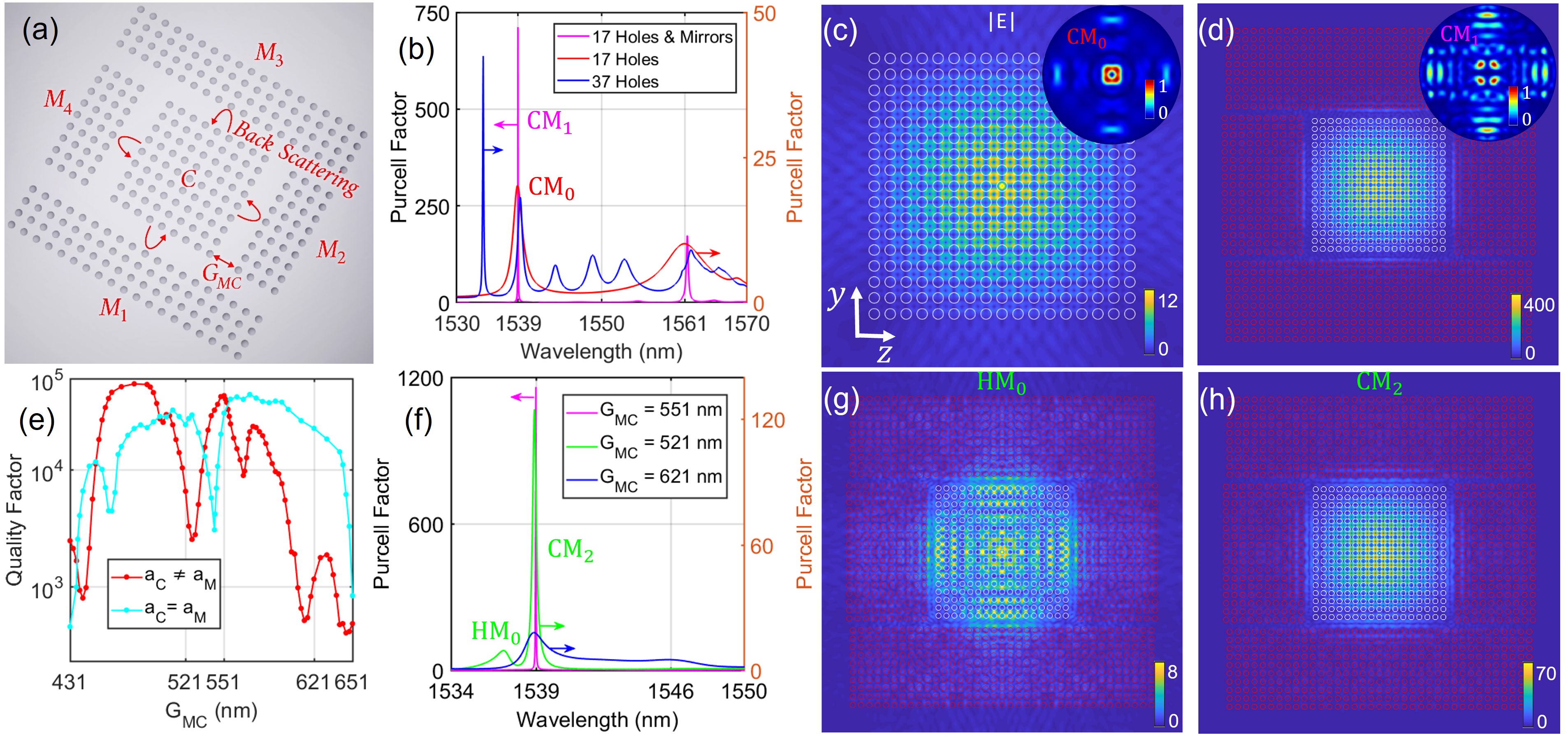}
\caption{\label{F1}
Unconventional localization via merging collective and hybrid modes.
(a) Supercavity with hole array (C) as cavity, surrounded by four mirror arrays (M$_{1,2,3,4}$) that backscatter in-plane leakage.
(b) Purcell factor spectrum for a magnetic dipole ($m_x$) at the cavity center. Tuning the cavity-mirror gap to $G_{\text{MC}} = 541$ nm enhances the factor over tenfold at CM$_1$.
(c,d) Near-field profiles of CM$_0$ and CM$_1$ showing strong enhancement from mirror-induced backscattering; insets show far-field patterns.
(e) $Q$-factor versus gap $G_{\text{MC}}$ for mismatched ($a_M = 552$ nm, $a_C = 529$ nm), and matched ($a_M = a_C = 529$ nm) lattice periods; both yield similar $Q$ enhancement.
(f) Purcell spectra at three gaps show $Q$ enhancement from HM$_0$–CM$_2$ coalescence.
(g,h) Near-field profiles of HM$_0$ and CM$_2$.}
\end{figure*}
{\it Resonant Mie-tronics: beyond the Anderson localization.---} 
Over the past four decades, research on light localization via Mie scattering has followed two main directions: (i) disorder-driven Anderson mechanisms~\cite{Yamilov2025anderson}, and (ii) optical coupling between Mie scatterers~\cite{Utyushev2021collective}. Early photonic crystal studies suggested that coupling hinders localization~\cite{Anderson1985question}. Contrary to this, we show that coupling is essential for efficient localization. More fundamentally, we demonstrate how long-range interactions and the positive energy of light preclude true Anderson localization.

The electromagnetic field inside a dielectric resonator can be expanded in regular electric ${\bf N}_{lm}$ and magnetic ${\bf M}_{lm}$ multipoles or equivalently as a superposition of plane waves~\cite{Devaney1974multipole,Hoang2014multipole}:
\begin{eqnarray}
{\bf E}({\bf r})&=& 
 \sum_{l=1}^{L}\sum_{m=-l}^{l}\left[\zeta_{lm}{\bf N}_{lm}(k{\bf r})+\eta_{lm}{\bf M}_{lm}(k{\bf r})\right] \label{E1} \\
 &=&\frac{ik}{2\pi}\int_0^{2\pi}\!d\beta\int_{0}^\pi\!d\alpha\,\sin\alpha\,\hat{E}(\hat s)e^{i{\bm k}\cdot{\bm r}}, \label{E2}
\end{eqnarray}
where $\zeta_{lm}$ and $\eta_{lm}$ are the multipolar coefficients and $\hat{E}(\hat s)$ is the amplitude of a plane wave with ${\bm k}=k(\sin\alpha\cos\beta,\sin\alpha\sin\beta,\cos\alpha)$. As $k$ is real and positive, all components are propagating—unlike the exponential localization in electronic Anderson localization.

For coupled resonators, multipoles interact via translational addition theorems~\cite{Chew1993efficient}. For instance, for two on-axis scatterers, an electric dipole (ED) ${\bf N}_{1;1}$ from one induces a magnetic dipole (MD) ${\bf M}_{1;1}$ in the other via the coupling coefficient:
\begin{equation}
B_{1;1}^{1;1} = \frac{\sqrt{3}}{2} e^{ikd} \left( \frac{i}{kd} - \frac{1}{(kd)^2} \right), \label{E3}
\end{equation}
where $d$ is the center-to-center spacing. 

Unlike Anderson theory, which considers only short-range interactions decaying faster than $1/d^3$~\cite{Anderson1958absence}, Eq.~\eqref{E3} exhibits a leading $1/d$ decay, highlighting long-range coupling as intrinsic to light localization. This analytic result complements full-wave simulations showing the absence of true Anderson localization in Mie systems~\cite{Yamilov2023anderson}. Moreover, such long-range interactions invalidate tight-binding approximations often used to describe photonic coupling~\cite{Rider2022advances,Hoang2024photonic}, even for high-$Q$ resonators~\cite{Yariv1999coupled,Hoang2017fano}.

{\it High-$Q$ Localization via Hybrid Super-Resonances.---} Mie-tronics provides a systematic framework for optimizing high-$Q$ resonances. We illustrate this with light localization in finite hole arrays [Fig.~\ref{F1}(a)]. The accompanying paper~\cite{Hoang2025symmetry} shows that arrays of spheres, squares, T-shapes, and holes all sustain the same collective MD resonance despite geometric differences. Among these, finite hole arrays confine light least effectively owing to strong in-plane leakage through the slab waveguide—yet redirecting this leakage toward the array center produces strong confinement via hybrid super-resonances.

Figure~\ref{F1}(b) compares Purcell spectra for: (1) a $17\times17$ hole-array cavity surrounded by four mirror arrays, (2) a stand-alone $17\times17$ array, and (3) a larger $37\times37$ array without mirrors. Configuration (1) [mode CM$_1$] enhances the Purcell factor by more than an order of magnitude compared with (2) and (3) [mode CM$_0$], an effect often attributed to a photonic bandgap from mismatched mirror periodicity~\cite{Chen2022observation}. Within Mie-tronics, however, this enhancement arises more directly from controlled resonance detuning between cavity and mirror arrays.

The number of unit cells determines the detuning between collective mirror and cavity modes, which governs the in-plane backscattering strength. Hence, even for matched lattice periods ($a_C=a_M$), tuning the mirror–cavity gap $G_{\text{MC}}$ can increase the $Q$ factor by two orders of magnitude [Fig.~\ref{F1}(e)]. Near- and far-field maps [Figs.~\ref{F1}(c,d)] show that backscattering amplifies near-field intensity by more than tenfold and enhances large-angle radiation—signs of reinforced in-plane feedback.

The mirrors also reshape vortex topology [insets of Figs.~\ref{F1}(c,d)], indicating strong coupling from resonance coalescence~\cite{Hoang2024photonic}. Purcell spectra for three gaps [Fig.~\ref{F1}(f)] show that at suboptimal gaps ($G_{\text{MC}}=521$ and $621$ nm), two modes appear: HM$_0$ and CM$_2$. CM$_2$, governed by the central array, stays nearly fixed in frequency, while HM$0$ redshifts with increasing $G_{\text{MC}}$, consistent with FP behavior. Field profiles [Figs.~\ref{F1}(g,h)] confirm their origins: HM$_0$ results from backreflection modified by Mie scattering, whereas CM$2$ is a collective mode similar to Fig.~\ref{F1}(c). At the optimal gap $G_{\text{MC}}=551$ nm, the two coalesce, with HM$_0$ constructively reinforcing CM$_2$.

\begin{figure}
\includegraphics[width = 6.5 cm]{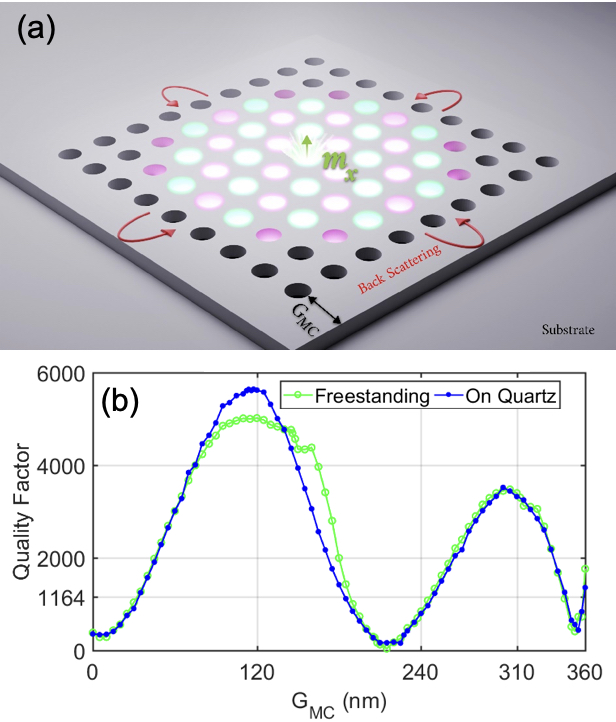}
\caption{\label{F2} (a) Schematic of a magnetic dipole ($m_x$) coupled to a hole array with four reflective edges. 
(b) $Q$-factor enhancement from edge-induced feedback, further improved by introducing a quartz substrate.}
\end{figure}

To isolate the feedback mechanism, we replace the mirror arrays in Fig.~\ref{F1}(a) with reflective edges [Fig.~\ref{F2}(a)]. These act as efficient FP mirrors~\cite{SuppMat} and support hybrid confinement. Figure~\ref{F2}(b) shows that edge reflections raise the $Q$ factor of the collective mode, though less effectively than mirror arrays: from $Q=1164$ in an infinite slab ($G_{\text{MC}}\to\infty$) to $\sim 6000$ at the optimal gap $G_{\text{MC}}=120$ nm. This contrast underscores the dominant role of collective, phase-matched backscattering in resonantly reinforcing localized modes.

Mechanical designs often require a quartz substrate, which further improves $Q$ near optimal configurations, indicating that vertical symmetry breaking can strengthen localization. Appendix~\ref{A1} shows that hybridizing FP and Mie (whispering-gallery) resonances yields additional hybrid modes that combine their advantages, enabling unconventional confinement regimes. While edge reflectors cannot reach the $Q$ of hole-array mirrors, they offer a clear advantage for integration by minimizing device footprint—a key feature of subwavelength Mie-tronics.

This focus on compactness naturally raises a question: for a fixed number of scatterers (e.g., a $17\times17$ hole array), can we reconfigure them into a smaller footprint while simultaneously enhancing both the Purcell and $Q$ factors? An answer emerges from an analogy with twisted bilayer graphene, where geometric reconfiguration and interference lead to extreme wave localization.

\begin{figure*}
\includegraphics[width = 1.8\columnwidth]{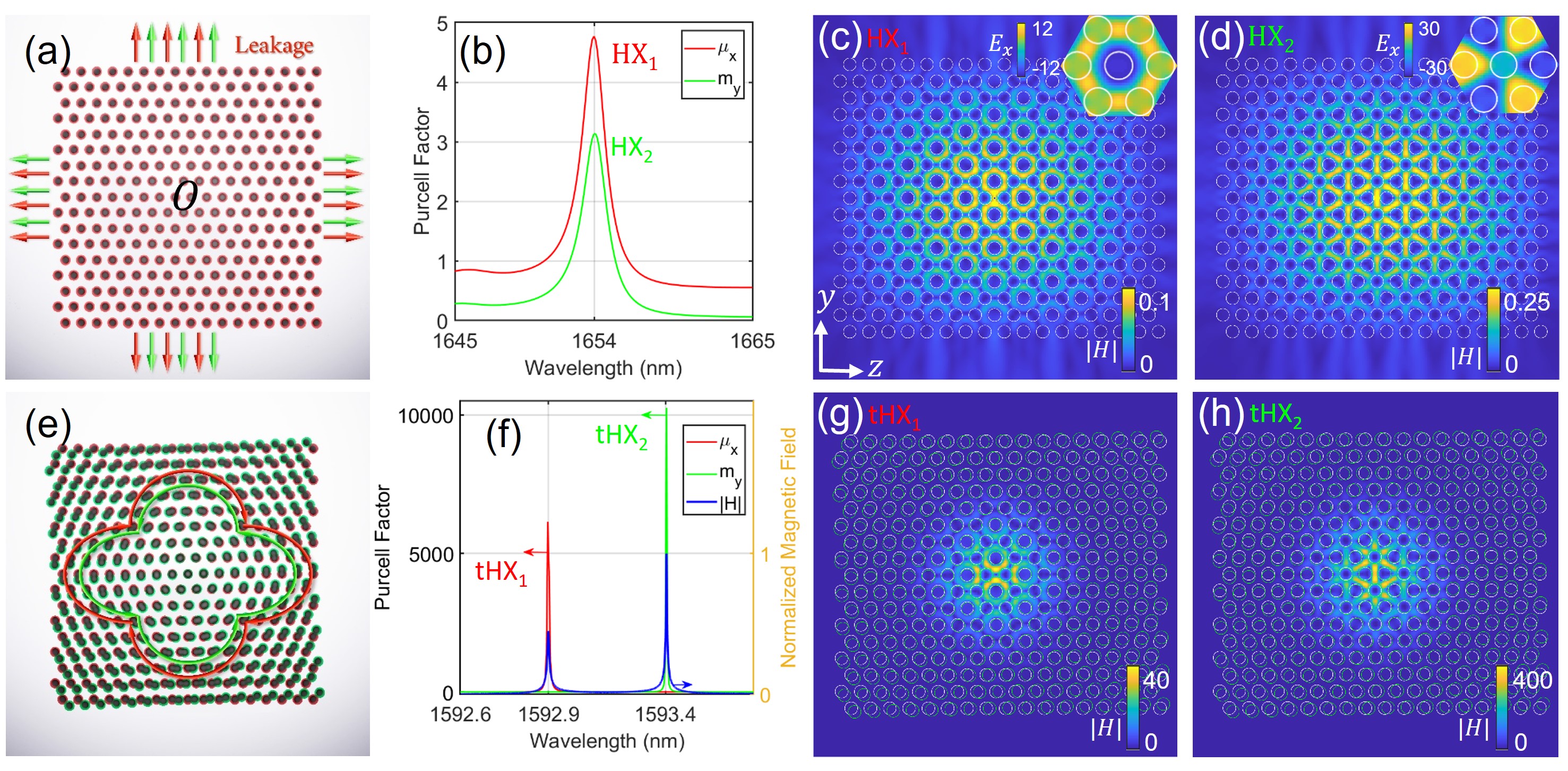}
\caption{\label{F3} Twisted Mie-tronics. (a) Hexagonal hole array, with red circles indicating air holes. (b) Purcell spectra at point $O$ showing two degenerate collective modes, HX$_1$ and HX$_2$. (c,d) Magnetic field profiles of HX$_1$ and HX$_2$, with insets showing their $E_x$ field symmetries. (e) Twisted hexagonal structure formed by overlaying a rotated copy (green). (f) Coupling of HX$_1$ and HX$_2$ at twist angle $\theta = 1.45^\circ$ results in twisted modes tHX$_1$ and tHX$_2$. (g,h) Near-field profiles of tHX$_1$ and tHX$_2$.}
\end{figure*}

{\it Localization in Twisted Mie-tronics via Strong Coupling.---}In twisted bilayer graphene, electrons localize at specific twist angles—so-called magic angles~\cite{Bistritzer2011moire}. History appears to repeat itself, as electronic magic angles have quickly inspired a parallel wave of interest in photonics~\cite{Mao2021magic, Raun2023gan}. Yet despite over a decade of intense investigation, the physical origin of magic angles remains debated~\cite{Tarnopolsky2019origin, Yu2023origin}. A major obstacle is the intractability of full-scale electronic structure calculations~\cite{Bistritzer2011moire}, leading to reliance on simplified tight-binding models, whose differing assumptions yield inconsistent predictions~\cite{Yu2023origin}.

Here we turn to Mie-tronics, where full-wave electromagnetic simulations can directly probe the photonic analogs of these effects. Owing to the shared wave nature of electrons and photons, the insights gained here may illuminate their electronic counterparts.

A natural starting point is to slightly twist the square hole array from Fig.~\ref{F1}(c). This twist leaves the central holes largely unchanged while significantly modifying the outer ones. One might expect the altered periphery to backscatter the in-plane leakage, thereby enhancing confinement. However, this strategy fails: the twist disrupts phase coherence among the interfering multipoles of the scatterers, weakening their collective resonance~\cite{Hoang2025collective}.

\begin{figure*}
\includegraphics[width = 1.8\columnwidth]{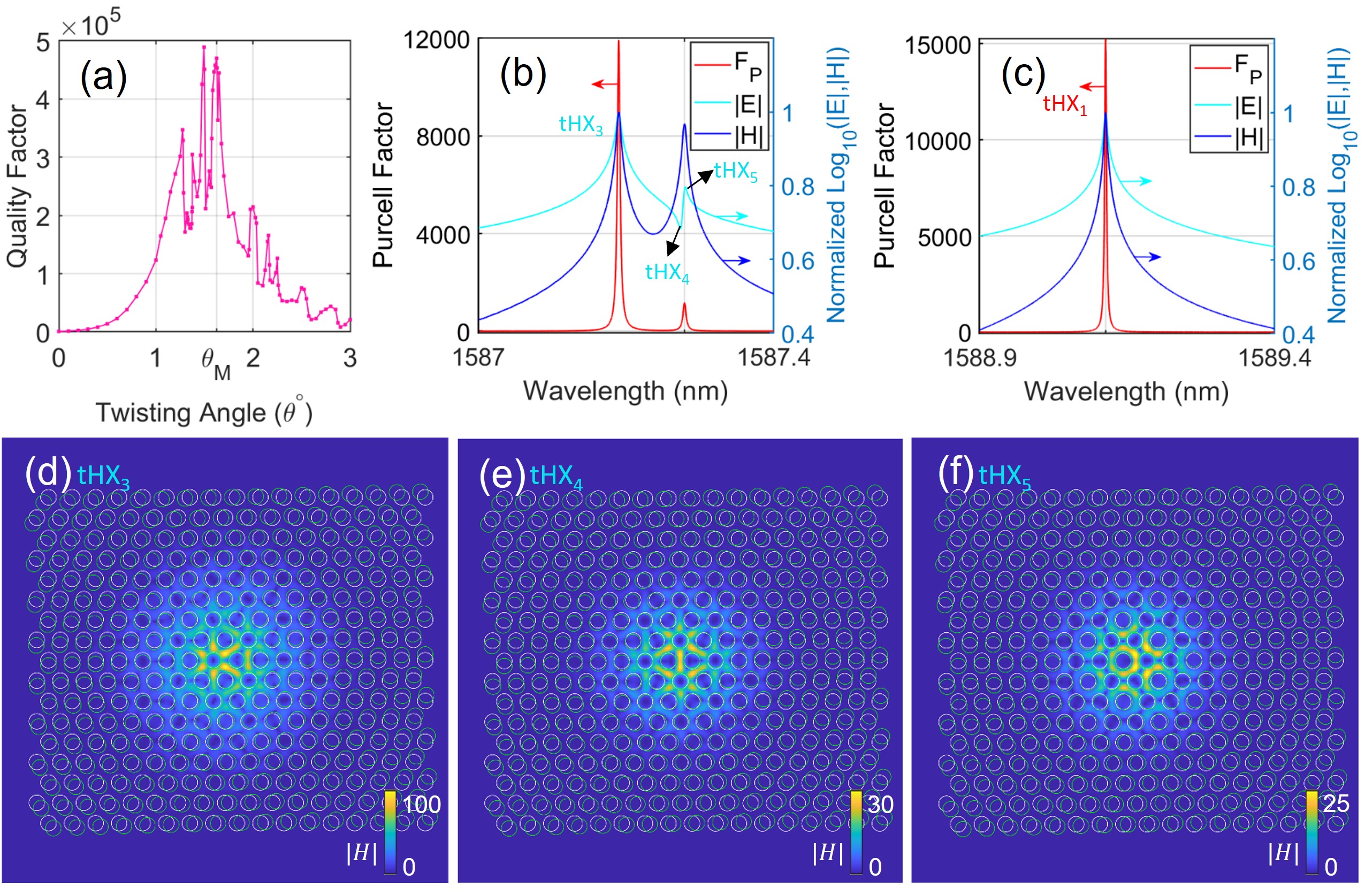}
\caption{\label{F4} Magic angles in Mie-tronics.
(a) $Q$ factor of the twisted mode tHX$_1$ versus twist angle $\theta$, showing multiple peaks—photonic analogs of electronic magic angles.
(b) Purcell spectrum at $\theta = 1.675^\circ$, with two peaks (tHX$_3$, tHX$_5$) and a trough (tHX$_4$).
(c) Spectrum at the magic angle $\theta_M = 1.6^\circ$, where $\mu_x$ excites only tHX$_1$.
(d–f) Magnetic field profiles of tHX$_3$, tHX$_4$, and tHX$_5$ from (b).}
\end{figure*}

This failure underscores the importance of lattice geometry. In graphene, the underlying lattice is hexagonal, not square. Likewise, rearranging the square array into a hexagonal lattice [Fig.~\ref{F3}(a)] and applying a small twist produces strongly localized modes [Figs.~\ref{F3},~\ref{F4}]. These modes originate from two degenerate collective resonances, HX$_1$ and HX$_2$ [Fig.~\ref{F3}(b)], with even and odd $E_x$ symmetries [Figs.~\ref{F3}(c,d)]. Because of these symmetry relations, a single dipole ($\mu_x$ or $m_y$) excites only one mode (HX$_1$ or HX$_2$), despite spatial and spectral overlap—an example of symmetry-enforced mode decoupling~\cite{Awai2007coupling}.

Although the moir\'e metastructure in Fig.~\ref{F3}(e) results from twisting arrays within a single silicon layer—distinct from stacked bilayers in graphene—it performs an analogous role: inducing interband coupling between symmetry-broken collective modes. Twisting the hexagonal array by $\theta = 1.45^\circ$ couples HX$_1$ and HX$_2$, yielding two hybrid modes, tHX$_1$ and tHX$_2$ [Fig.~\ref{F3}(f)]. The magnetic spectrum at point $O$ under $m_x$ excitation confirms their coupling, while the field profiles [Figs.~\ref{F3}(g,h)] reveal two to three orders of magnitude of enhancement and strong central localization—hallmarks of interband interference suppressing in-plane leakage. Similar signatures appear experimentally as van Hove singularities in twisted bilayer graphene~\cite{Li2010observation} and in dual-mode lasing of photonic moir\'e structures~\cite{Raun2023gan}.

{\it Photonic magic angles.---}To identify magic angles, we plot the $Q$ factor of tHX$_1$ versus twist angle $\theta$ [Fig.~\ref{F4}(a)]. For $\theta>1^\circ$, multiple peaks emerge, with the maximum $Q$ exceeding that in Fig.~\ref{F1} by more than fivefold despite occupying only 18$\%$ of the $37\times 37$ footprint. For $\theta > 3^\circ$, the $Q$ factor drops below $10^4$ as inner cells detune, offsetting the interband enhancement.

Fluctuations near the $Q$ peaks reflect alternating constructive and destructive interference within the moir\'e structure. At a suboptimal twist $\theta = 1.675^\circ$ [Fig.~\ref{F4}(b)], two modes—tHX$_3$ and tHX$_5$—coexist with a suppressed mode tHX$_4$. Mode tHX$_5$, derived from HX$_1$, destructively interferes with tHX$_4$ from HX$_2$, lowering both $Q$ and Purcell enhancement and distorting the field [Fig.~\ref{F4}(f)]. The emergence of tHX$_3$ [Fig.~\ref{F4}(d)], also from HX$_2$, coincides with the suppression of tHX$_4$ [Fig.~\ref{F4}(e)], confirming strong interband coupling.

At the magic angle $\theta_M = 1.6^\circ$ [Fig.~\ref{F4}(c)], both tHX$_3$ and tHX$_4$ vanish, leaving only tHX$_1$ coupled to the ED excitation $\mu_x$. Twisted HX$_2$ modes persist, but are accessible only via MD excitation ($m_y$). This behavior underscores the distinct nature of interband coupling—qualitatively different from intraband interactions~\cite{Hoang2024photonic} and from the hybrid supercavity modes in Fig.~\ref{F1}. Here, magic angles arise from alternating spectral crossings, not mode coalescence.

Despite the strong localization, the electromagnetic coupling remains long-ranged, extending across several unit cells within each moir\'e supercell. Interband coupling suppresses outer-cell fields, effectively decoupling neighboring supercells and producing flat photonic bands~\cite{SuppMat}. Yet, because the positive energy of light enforces radiation, flatbands do not yield infinite $Q$ or perfect confinement. Unlike periodic arrays~\cite{Hoang2025collective}, enlarging the system does not lead to divergent $Q$, even when the dispersion appears flat. This radiative, long-range nature explains why disorder typically weakens—rather than enhances—localization (Appendix~\ref{A2}).

Although the radiative and long-range character of Mie scattering was recognized in early theories~\cite{John1991localization,Anderson1985question}, the implications for Anderson localization were not fully resolved, largely due to computational limitations of the time—similar to those faced in early electronic-structure studies of moir\'e materials~\cite{Bistritzer2011moire}. The present results clarify that these intrinsic electromagnetic properties—radiation and long-range coupling—fundamentally constrain optical localization, preventing it from reaching the fully trapped regime envisioned in earlier models.

Leveraging modern computational tools and the analytical framework of Mie-tronics, we find that the strongest localization arises from intra- and interband coupling. Even these modes—combining electric hotspots with flatband resonances—remain sub-exponentially localized~\cite{Hoang2024photonic}. Mie-tronics thus provides a unified framework for understanding the electromagnetic limits of light localization—showing that only by embracing its inherently radiative and long-range nature can one achieve optimal confinement.

{\it Conclusion.---}We have developed Mie-tronics as a general and unifying framework for understanding and optimizing light localization in subwavelength systems. By combining the strengths of Fabry-P\'erot and collective Mie resonances, we demonstrated how hybrid super-resonances can substantially enhance both the $Q$ and Purcell factors. We further elucidated the origin of photonic magic angles, showing that twist-induced coupling between degenerate collective modes in moir\'e structures yields strong confinement analogous to flatband localization in twisted bilayer graphene.

Contrary to conventional wisdom, we find that symmetry breaking and strong coupling—rather than disorder—enable the most efficient pathways to light localization. Together with the accompanying paper~\cite{Hoang2025symmetry}, these findings position Mie-tronics as a foundation for designing next-generation photonic architectures, with implications extending beyond optics to other waves.

\begin{acknowledgments}
This research is supported by the National Research Foundation, Singapore, and A*STAR under its Frontier Competitive Research Programme (NRF-F-CRP-2024-0009), the Australian Research Council (Grant No. DP210101292), and International Technology Center Indo-Pacific (ITC IPAC) via Army Research Office (Contract FA520923C0023). D.L. acknowledges a support from the Ministry of Education of Singapore under its  SUTD Kickstarter Initiative (Grant No. SKI 20210501). J. Ji acknowledges the funding received from the PhotonDelta National Growth Fund programme.
\end{acknowledgments}

\bibliography{Reference_PRL}
\clearpage

\appendix
\section{Hybridization of Fabry-P\'erot and Mie Resonances} \label{A1}
\begin{figure}[htbp]
\includegraphics[width = 0.9\columnwidth]{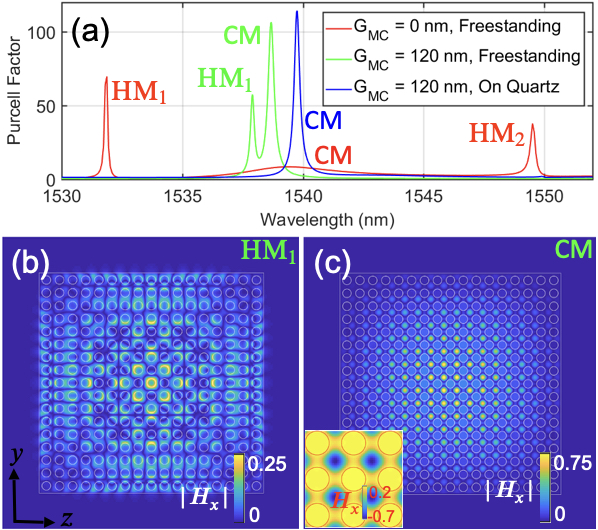}
\caption{\label{A1F1} Merging of hybrid and collective modes, and impact of a quartz substrate.
(a) Purcell factor spectra showing the merging of the hybrid mode (HM$_1$) with the collective mode (CM) as we tune the gap size $G_{\text{MC}}$ from 0 to 120~nm. Adding a quartz substrate slightly enhances the CM while suppressing HM$_1$. (b,c) Near-field profiles of HM$_1$ and CM, respectively. The inset in (c) shows the imaginary part of $H_x$, highlighting opposite magnetic field directions in hole-centered and silicon-centered unit cells.}
\end{figure}

Figure~\ref{A1F1} shows how resonance merging contributes to the $Q$ enhancement in Fig.~\ref{F2}(b). At $G_{\text{MC}} = 0$ nm, two hybrid modes (HM$_1$, HM$_2$) emerge. Compared with FP modes in the absence of holes~\cite{SuppMat}, these exhibit lower mode density and higher Purcell factors—features desirable for single-mode nanolasers.

As $G_{\text{MC}}$ increases, the hybrid modes redshift linearly due to their FP character, while the collective mode (CM) remains fixed near 1539 nm. At $G_{\text{MC}} = 120$ nm, the CM reaches maximum enhancement, boosted by merging with the auxiliary HM$_1$. A quartz substrate suppresses HM$_1$ but slightly enhances CM, consistent with the symmetry-breaking effects discussed in the accompanying paper~\cite{Hoang2025symmetry}.
This shows the advantage of the antibonding collective mode over its bonding counterpart.

Figures~\ref{A1F1}(b,c) compare the near-field profiles of HM$_1$ and CM.
The inset in Fig.~\ref{A1F1}(c) shows the imaginary magnetic field concentrated inside both hole- and silicon-centered unit cells, with opposite signs along $\pm x$.
The hole components correspond to Mie void modes~\cite{Hentschel2023dielectric}, distinguishing this lattice from sphere, square, and T-shaped arrays~\cite{Hoang2025symmetry}. Hybridization between Mie modes in silicon-centered cells and void modes in air holes arises from the silicon slab acting as a FP cavity, enabling unconventional localization not found in other geometries.
The distinct field profile of HM$_1$ [Fig.~\ref{A1F1}(b)] further differentiates hole arrays from reflective-edge designs~\cite{SuppMat}.

\begin{figure}[htbp]
\includegraphics[width = 0.9\columnwidth]{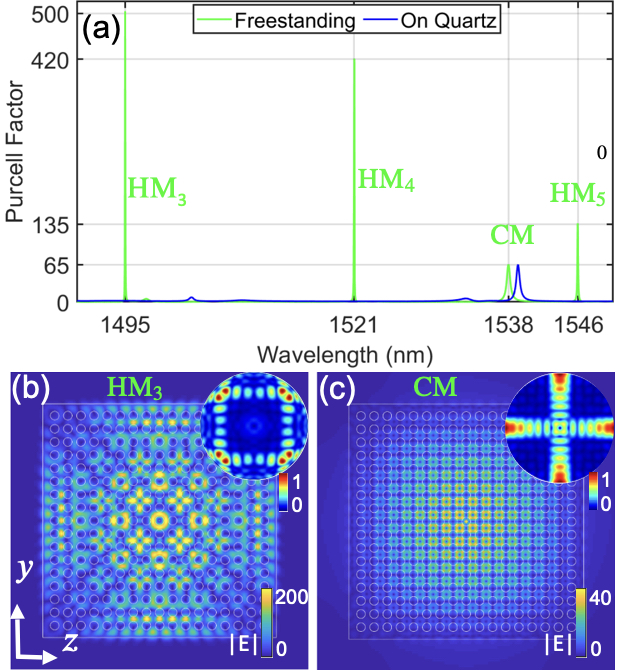}
\caption{\label{A1F2} (a) Purcell spectra for a freestanding metastructure, showing three hybrid modes (HM$_{3,4,5}$) and a collective mode (CM); only CM survives with a substrate. 
(b,c) Near-field profiles of HM$_3$ and CM, with insets showing their far-field radiation patterns.}
\end{figure}

At larger gaps ($G_{\text{MC}} = 310$~nm), Fig.~\ref{A1F2}(a) reveals three additional hybrid modes (HM$_3$–HM$_5$) exhibiting Purcell enhancements several times higher than the CM. Although spectrally resembling FP resonances, their field and angular distributions indicate a more complex origin. The far-field insets of Figs.~\ref{A1F2}(b,c) show that vortex-like beams persist but are no longer dominant; instead, radiation spreads into multiple angular channels, reflecting interference among distinct localization mechanisms.

Comparing the electric-field profiles of HM$_3$ and CM [Figs.~\ref{A1F2}(b,c)] illustrates this interplay. The HM$_3$ pattern cannot be ascribed to a single mechanism—FP reflection or collective Mie resonance alone—but results from their coherent coupling. In contrast, CM remains primarily governed by collective MD resonance, though its multi-angle radiation pattern evidences FP feedback.

These results reveal a general principle: efficient light localization emerges from the constructive interference of multiple feedback mechanisms. FP reflections, substrate-induced confinement, and collective resonances act cooperatively to produce high-$Q$ hybrid modes—achieving confinement levels unattainable by any single mechanism.

\begin{figure*}[htbp]
\includegraphics[width=1.8\columnwidth]{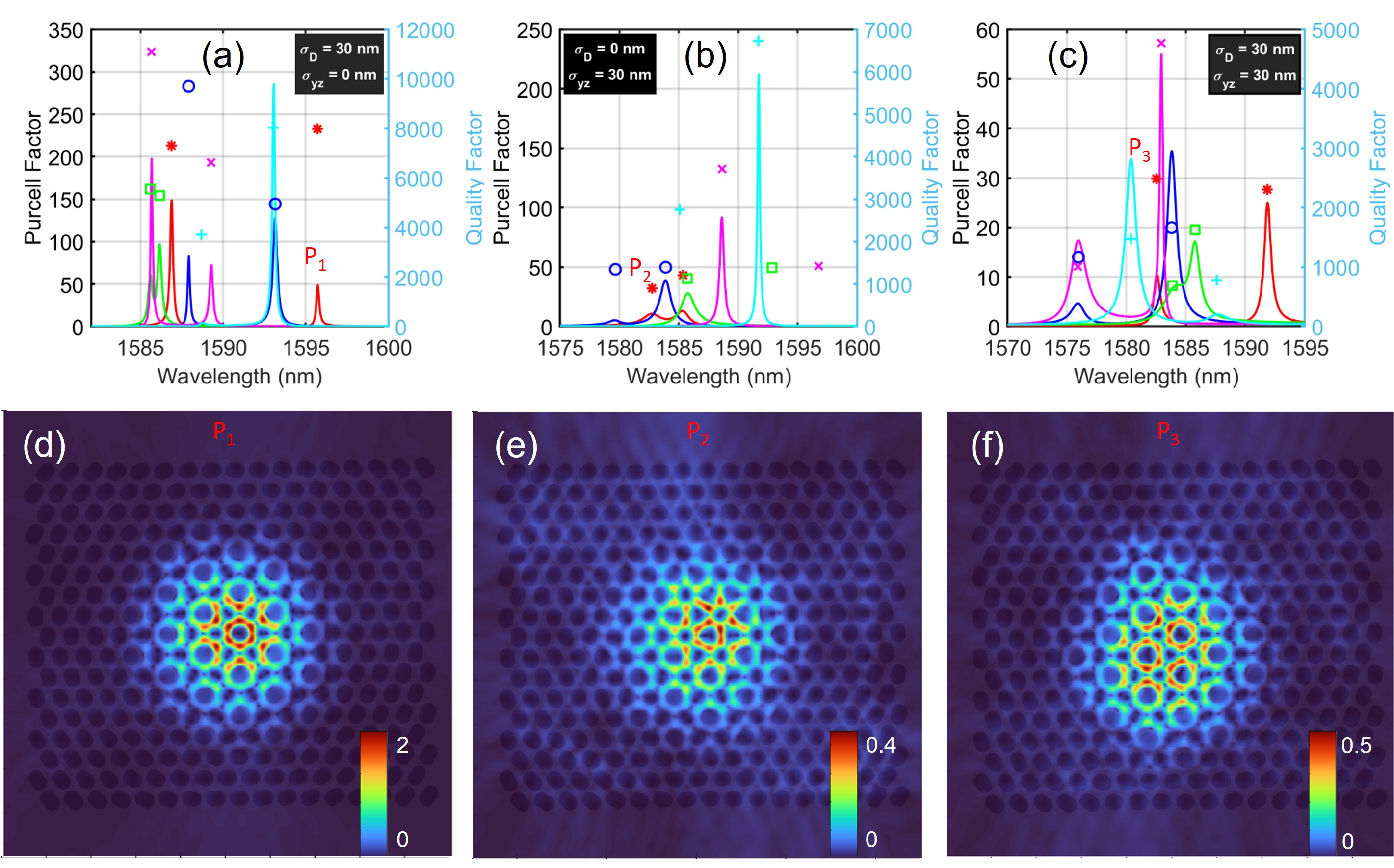}
\caption{\label{A2F1} Impact of diameter and positional disorder on the resonance at the magic angle $\theta = 1.5^\circ$.
(a) Diameter disorder ($\sigma_D = 30$~nm) reduces the Purcell and $Q$ factors by more than tenfold and splits the collective resonances.
(b) Positional disorder ($\sigma_{yz} = 30$~nm) causes even stronger degradation.
(c) Combined disorder further reduces both metrics, compounding (a) and (b).
(d–f) Field profiles of the disordered modes from (a–c) show increased in-plane leakage and weakened localization.}
\end{figure*}
\section{Disorder-Induced Degradation of Light Localization in Moir\'e Cavities}\label{A2}

To quantify the effects of disorder on localization in moir\'e cavities, we introduced controlled variations in hole diameter ($\sigma_D$) and position ($\sigma_{yz}$). Here, $\sigma_D=30$~nm corresponds to pseudorandom diameters $D_h$ uniformly distributed within $[335,365]$~nm, while positional disorder applies similarly along $y$ and $z$~\cite{Hoang2025collective}.
Fifteen disordered realizations were simulated, summarized in Fig.~\ref{A2F1}.

In the Purcell spectra [Fig.~\ref{A2F1}(a–c)], disorder progressively weakens and splits the collective resonances. In some cases, the resonances disappear from the Purcell spectrum but remain identifiable in the magnetic field spectrum at the cavity center; their $Q$ factors are then extracted from the temporal decay of the field.
Across all cases, disorder consistently reduces both the $Q$ factor and spatial confinement.

Increasing disorder from Fig.~\ref{A2F1}(a) to Fig.~\ref{A2F1}(c) weakens constituent resonances and disrupts the interference necessary for forming high-$Q$ twisted modes.
Spectral peaks shift, broaden, and separate, diminishing overlap and coherent backscattering. As a result, radiation losses—especially in-plane—rise significantly, as confirmed by the near-field maps in Figs.~\ref{A2F1}(d–f).

Localized modes in photonic structures are known to coexist with a diffusive background that decays algebraically with distance~\cite{Lagendijk1996resonant}. Anderson localization requires suppression of this diffusive component. In contrast, our results show that in Mie-tronic systems, long-range radiative coupling persists under disorder, enhancing diffusion rather than quenching it.

Moreover, increased spectral splitting in disordered structures does not indicate stronger coupling; it instead reflects weakened interference due to random spatial and spectral perturbations. This counterintuitive behavior underscores the delicate interplay among interference, radiation, and localization in Mie-tronic systems—and highlights that disorder, rather than confining light, fundamentally degrades it.

\clearpage
\onecolumngrid 
\begin{center}
\textbf{\large Supplemental Material for\\
``Unconventional Localization of Light with Mie-tronics''}
\end{center}
\medskip

\setcounter{equation}{0}
\setcounter{figure}{0}
\setcounter{table}{0}
\setcounter{section}{0}
\renewcommand{\theequation}{S\arabic{equation}}
\renewcommand{\thefigure}{S\arabic{figure}}
\renewcommand{\thetable}{S\arabic{table}}
\renewcommand{\thesection}{S\arabic{section}}

\twocolumngrid

\section{Fabry-P\'erot Feedback: Mietronic Hole Arrays vs. Conventional Mirrors} \label{S1}
\begin{figure}[htbp]
\includegraphics[width = 8 cm]{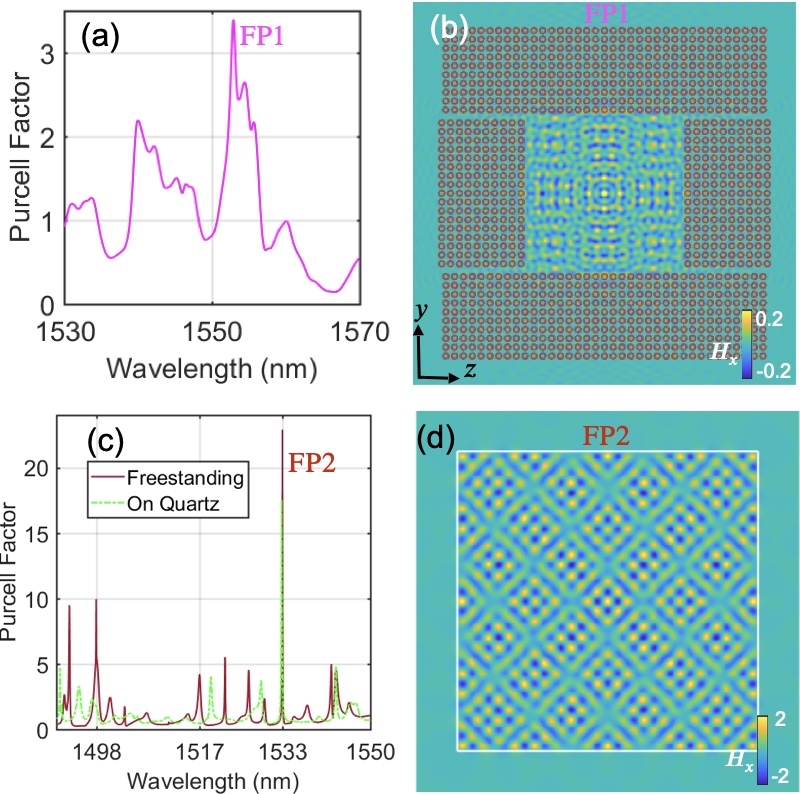}
\caption{\label{S1F1}Light confinement by photonic-crystal mirrors and reflective edges.
(a) Spectral profile of the $m_x$ magnetic dipole located at the center of the cavity, formed by removing the $17\times17$ hole array from the cavity structure shown in Fig. 1 of the main text, with $G_{MC} = 551$ nm. (b) Magnetic $H_x$ field distribution corresponding to the highest peak of the Purcell factor shown in (a). (c) Spectral profiles for both the freestanding cavity and the case with a quartz substrate, with edge sizes corresponding to $G_{MC} = 310$ nm in Fig. 4 of the main text. (d) Magnetic $H_x$ field distribution for the pronounced Fabry-P\'erot mode marked in (c).}
\end{figure}
In this section, we investigate how hole arrays function as photonic crystal mirrors and compare them to Fabry-P\'erot (FP) mirrors formed by four reflective edges. While both types of mirrors confine light within their central region and support FP modes, their underlying mechanisms differ due to the collective nature of the hole arrays.

Conventional edge mirrors backscatter light uniformly across all wavelengths, facilitating the formation of standing wave patterns characteristic of FP modes. In contrast, hole array mirrors exhibit wavelength-dependent backscattering, governed by their interaction with collective Mie modes. This interaction results in distinct optical properties, including enhanced mode selectivity and modified spectral responses.

Figures~\ref{S1F1}(a) and \ref{S1F1}(b) illustrate the interaction between a MD emitter ($m_x$) and an FP cavity formed by four hole arrays. The bandwidth of the hole-based photonic crystal extends from 1536 nm to beyond 1715 nm, encompassing the peak of the Purcell factor shown in Fig.~\ref{S1F1}(a). This observation presents a limit of explanations based solely on the photonic crystal framework. Within this framework, the FP cavity can be regarded as a defect, similar to the well-known $L_3$ photonic crystal cavity. However, in this case, light confinement does not rely on photonic band gaps since the prominent peak in Fig.~\ref{S1F1}(a) lies within the band-pass region. In contrast, Mie-tronics provides a much clearer picture for understanding the underlying mechanism.

In Mie-tronics, hole arrays collectively act as backscatterers, determining which wavelengths are efficiently reflected based on the number of holes rather than photonic band gaps. These hole arrays also allow partial transmission and radiation leakage into the surrounding free space. As a result, the Purcell peak observed in Fig.~\ref{S1F1}(a) remains relatively modest, consistent with the moderate field confinement observed in Fig.~\ref{S1F1}(b). Nonetheless, the field distribution retains the defining features of a typical FP mode, with comparable peak intensities near both the center and the edges of the cavity.

Figures~\ref{S1F1}(c) and \ref{S1F1}(d) show the interaction between an FP cavity formed by four reflective edges and a magnetic dipole $m_x$ placed at its center. The spectral responses in Fig.~\ref{S1F1}(c) exhibit characteristics typical of FP modes, including evenly spaced resonances and high $Q$ factors. The relatively low peak Purcell factor results from the standing wave being uniformly distributed throughout the FP cavity, as shown in Fig.~\ref{S1F1}(d). Despite the high $Q$ factor of the FP mode ($>10^4$), the relatively small mode volume limits the achievable Purcell enhancement.

Notably, the main text shows that introducing a $17 \times 17$ hole array at the center of the FP cavity and fine-tuning the structure significantly increases both the Purcell factor and the $Q$ factor for the hole-based cavity (Fig.~1) compared to the edge-based cavity (Fig.~4), highlighting the collective nature of the hole array.

\section{Origin of Flatbands in Twisted Moir\'e Metastructure} \label{S2}
This section further investigates interband coupling in twisted moir\'e metastructures and its impact on light localization and flatband formation.

\begin{figure}[htbp]
\includegraphics[width = 8 cm]{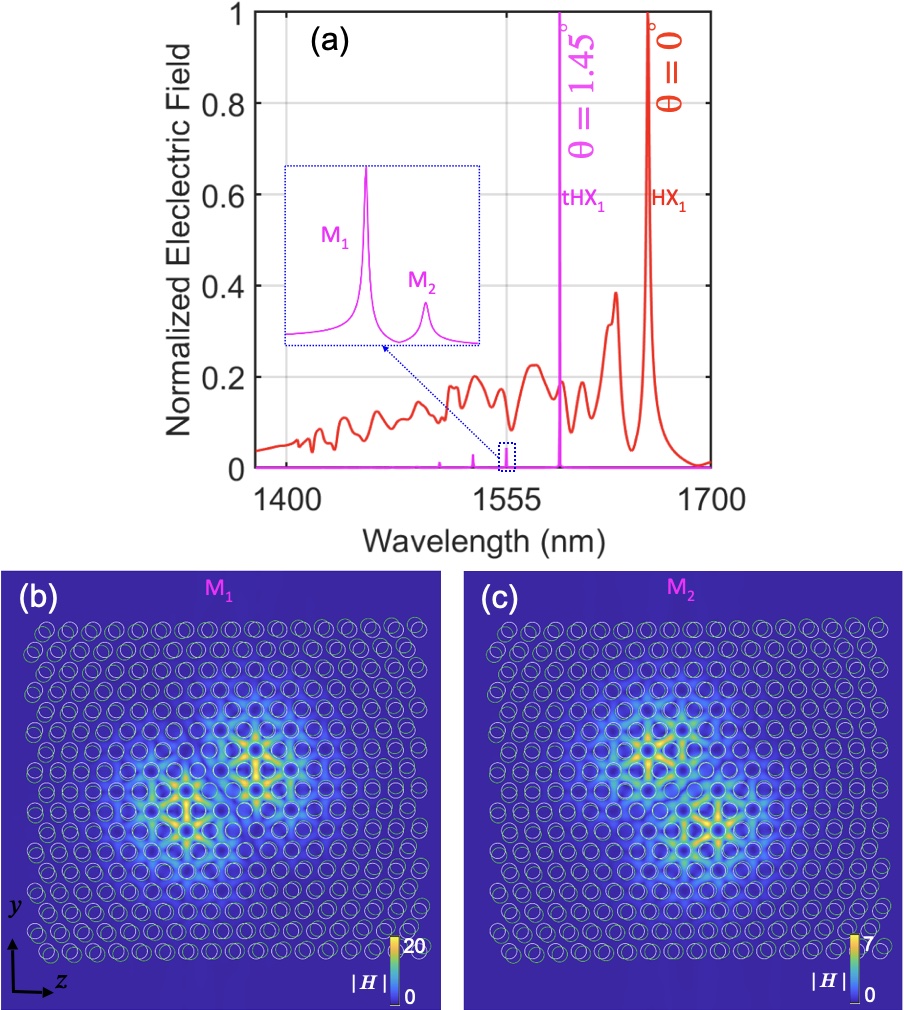}
\caption{\label{S2F1}Weaker collective resonances in the twisted metastructure. (a) Normalized electric field spectra of the excited modes in the hexagonal hole array ($\theta = 0^\circ$) and its twisted counterpart ($\theta = 1.45^\circ$), with the dipole source $\mu_x$ at the center. The inset highlights two resonant modes around 1555 nm, labeled M$_1$ and M$_2$, which are weaker than tHX$_1$. (b) and (c) Magnetic field distributions of M$_1$ and M$_2$, showing their common origin from band HX$_2$. In contrast, the main text discusses HX$_1$ and its twisted version tHX$_1$, which originate from band HX$_1$. This indicates that in the twisted metastructure, both bands strongly interact with the dipole source.}
\end{figure}

\begin{figure}[htbp]
\includegraphics[width = 8 cm]{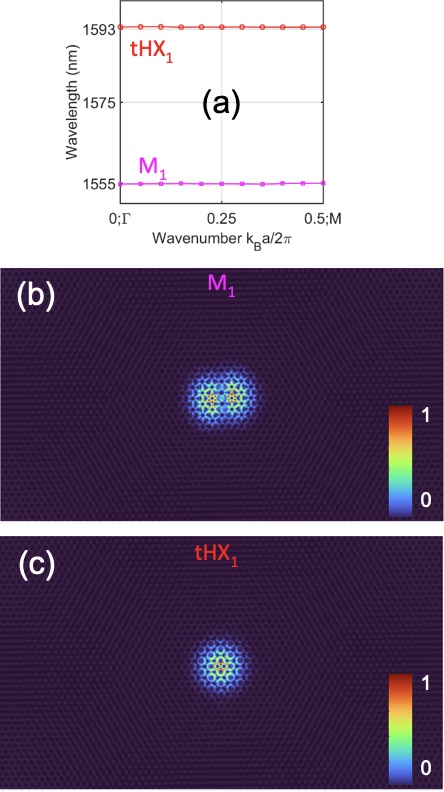}
\caption{\label{S2F2}Origin of moir\'e flatbands. (a) Two flatbands obtained by applying Bloch boundary conditions to the moir\'e unit cell corresponding to a twist angle of $\theta = 1.45^\circ$. (b) and (c) Magnetic field distributions of the eigenmodes associated with the flatbands in (a) reveal their tightly confined nature. This strong spatial confinement, together with the positive energy of light, renders the moir\'e modes insensitive to the imposed Bloch boundary conditions, giving rise to superflatbands.}
\end{figure}
Figure~\ref{S2F1}(a) shows the full spectral range of band HX$_1$ in the hexagonal hole metastructure discussed in the main text. In the untwisted structure ($\theta = 0^\circ$), this band is spectrally broad and dominated by the collective band-edge mode HX$_1$. Introducing a twist ($\theta = 1.45^\circ$) induces macroscopic degeneracy, compressing this broad band into a few distinct, extremely narrow spectral peaks. While HX$_1$ and its twisted counterpart tHX$_1$ are discussed in the main text, the inset in Fig.~\ref{S2F1}(a) reveals two additional modes, labeled M$_1$ and M$_2$. Interestingly, the magnetic field distributions of M$_1$ and M$_2$ in Figs. \ref{S2F1}(b) and \ref{S2F1}(c) indicate that they originate from band HX$_2$. This suggests that, unlike in the untwisted metastructure, the dipole source in the twisted metastructure strongly couples to both bands. Furthermore, the mode profiles of M$_1$ and M$_2$ clearly highlight their second-order collective nature, characterized by a null magnetic field at the cavity center.

Although moir\'e structures form true crystals only at specific discrete twist angles, both theoretical and experimental studies have shown that quasi-crystal moir\'e structures exhibit similar phenomena to their crystalline counterparts. Figure~\ref{S2F2} explains this observation from the perspective of Mie-tronics. The two collective modes are strongly localized at the cavity center, and most of the light leaks out in the out-of-plane direction. This effectively decouples the localized modes in neighboring moir\'e supercells, rendering the eigenmodes insensitive to Bloch boundary conditions. Both the flatband wavelengths in Fig.~\ref{S2F2}(a) and the field distributions in Figs.~\ref{S2F2}(b)-(c) indicate that these flatbands originate from the twisted collective resonances shown in Fig.~\ref{S2F1}(a).

\end{document}